\documentclass[letterpaper, 10 pt, conference]{ieeeconf}  % Comment this line out
\IEEEoverridecommandlockouts                              % This command is only
\usepackage{graphicx}      % include this line if your document contains figures
\usepackage{cite}

\usepackage{epsfig} % for postscript graphics files
\usepackage{mathptmx} % assumes new font selection scheme installed
\usepackage{times} % assumes new font selection scheme installed
\usepackage{amsmath} % assumes amsmath package installed
\usepackage{amssymb}  % assumes amsmath package installed
\usepackage{cancel}
\usepackage{gensymb}
\usepackage{amsfonts}
\usepackage{booktabs,tabularx}

\newcommand\numberthis{\addtocounter{equation}{1}\tag{\theequation}}

\title{\LARGE \bf
Complex order control for improved loop-shaping in precision positioning
}

\author{Niranjan Saikumar, Duarte Val\'erio and S. Hassan HosseinNia% <-this % stops a space
\thanks{This work was supported by NWO, through OTP	TTW project \#16335, by FCT, through IDMEC, under LAETA, project UID/EMS/50022/2019, and grant SFRH/BSAB/142920/2018 attributed to the second author.}% <-this % stops a space
\thanks{Niranjan Saikumar and S. Hassan HosseinNia are with Faculty of 3ME, TU Delft, The Netherlands
        {\tt\small n.saikumar@tudelft.nl, s.h.hosseinniakani@tudelft.nl}}%
\thanks{Duarte Val\'erio is with IDMEC, Instituto Superior T\'ecnico, Universidade de Lisboa, Lisboa, Portugal
        {\tt\small duarte.valerio@tecnico.ulisboa.pt}}%
}

\begin{document}

\maketitle
\thispagestyle{empty}
\pagestyle{empty}

%%%%%%%%%%%%%%%%%%%%%%%%%%%%%%%%%%%%%%%%%%%%%%%%%%%%%%%%%%%%%%%%%%%%%%%%%%%%%%%%
\begin{abstract}

This paper presents a complex order filter developed and subsequently integrated into a PID-based controller design. The nonlinear filter is designed with reset elements to have describing function based frequency response similar to that of a linear (practically non-implementable) complex order filter. This allows for a design which has a negative gain slope and a corresponding positive phase slope as desired from a loop-shaping controller-design perspective. This approach enables improvement in precision tracking without compromising the bandwidth or stability requirements. The proposed designs are tested on a planar precision positioning stage and performance compared with PID and other state-of-the-art reset based controllers to showcase the advantages of this filter.

\end{abstract}

%%%%%%%%%%%%%%%%%%%%%%%%%%%%%%%%%%%%%%%%%%%%%%%%%%%%%%%%%%%%%%%%%%%%%%%%%%%%%%%%
\section{INTRODUCTION}

In view of the increasing demands on speed and accuracy from the precision industry, linear controllers are proving to be incapable of meeting the requirements considering their inherent limitations namely - Bode's gain-phase relationship and waterbed effect \cite{aastrom2000limitations,freudenberg2000survey,MSDbook}. The unique relationship between gain and phase of linear systems creates a trade-off between tracking and steady-state precision on one side, and bandwidth and stability margins on the other. In fact, from the perspective of loop-shaping which is the industry standard technique for controller design in the frequency domain, this need to move away from the gain-phase relationship was recognized in complex order derivatives (a sub-set of non-integer order calculus) used in control as in the third generation CRONE technique \cite{oustaloup2000frequency,sabatier2015fractional}.

However, such a derivative - which can potentially provide negative gain slope with a corresponding positive, increasing phase - is still linear and unfortunately practically unimplementable; and existing methods of approximation in a frequency range of interest result in unstable poles, non-minimum phase zeros or poor approximation of gain behaviour among other issues \cite{oustaloup1999commande,moghadam2018tuning}. The impossible nature of achieving complex order behaviour in linear systems, and the desire and necessity to overcome these limitations have led towards a growing interest in nonlinear control methods with the caveat that these be industry compatible in design and implementation. One such interesting concept is reset control, which is often referenced using the Clegg integrator \cite{Clegg1958,chait2002horowitz}. In this regard, this paper presents the approximation of complex order control, hereafter referred to as CLOC, utilizing the nonlinearity of reset elements to obtain an equivalence in the frequency domain behaviour as analysed using describing function.

The concept of reset control was introduced through the Clegg integrator which has its state reset to zero when its input has a zero crossing \cite{Clegg1958}. Describing function analysis shows similar magnitude behaviour, but with reduced phase lag of $38.1^\circ$ compared to its linear counterpart. This has been extended in literature to include first-order reset element (FORE) \cite{zaccarian2005first}, second-order reset element (SORE) \cite{Hazeleger2016} and even fractional-order (intended for all non-integer real numbers) reset elements (FrORE) \cite{Saikumar2017} including generalization \cite{saikumar2018constant} to take advantage of this phase behaviour. However, most works in literature limit the potential of reset by utilizing it mainly as part of the integrator and/or using the reduced phase lag property. Recently, a novel reset filter termed `Constant-gain Lead-phase' (CgLp) was designed combining FORE or SORE with its corresponding linear lead filter \cite{arun2018,saikumar2018constant,saikumar2019resetting}. The describing function of CgLp yields a constant unity gain with a corresponding phase lead in the designed frequency range of interest enabling significant improvement in open-loop shape. Although the analysis and designs are based on describing function without considering higher order harmonics, significant improvements in tracking and steady-state precision are reported.

In continuation with this trend, it is logical that the next step is in the direction of obtaining complex order derivative behaviour such that phase lead with positive slope can be achieved with a negative gain slope in the desired frequency range. This will further enable improvements in open-loop shape without compromising stability margins. Hence, this paper presents the describing function based approximation of complex order filter using reset elements and the subsequent integration into PID-framework to obtain CLOC. The designs and expected improvement in performance are validated on a precision planar positioning stage since higher-order harmonics are neglected during design. As such, the necessary preliminaries are provided in Section \ref{Preliminaries}, followed by the design of complex order filters in Section \ref{cloc}. The integration of complex order filters to obtain CLOC for the precision positioning stage is explained in Section \ref{experimental}, and the tracking and steady-state precision results obtained are compared with PID and CgLp based PID controllers and analysed in Section \ref{results} followed by conclusions in Section \ref{conclusions}.

%%%%%%%%%%%%%%%%%%%%%%%%%%%%%%%%%%%%%%%%%%%%%%%%%%%%%%%%%%%%%%%%%%%%%%%%%%%%%%%%
\section{PRELIMINARIES}
\label{Preliminaries}

\subsection{Complex order derivative and control}

Complex order derivatives can be defined in several ways and are denoted by the operator $D^{\alpha+j\beta}$ \cite{samko1993,podlubny1999,valerio2013}. Since our main concern and aim is for approximation of the behaviour in frequency domain, the focus is on the complex power of Laplace variable $s$ as:
\begin{equation}
\mathcal{L} \left[ D^{\alpha+j\beta} f(t) \right] = s^{\alpha+j\beta} \mathcal{L}\left[f(t)\right]
\end{equation}

\begin{figure}
	\centering
	\includegraphics[trim = {1cm 0 1cm 0}, width=\linewidth]{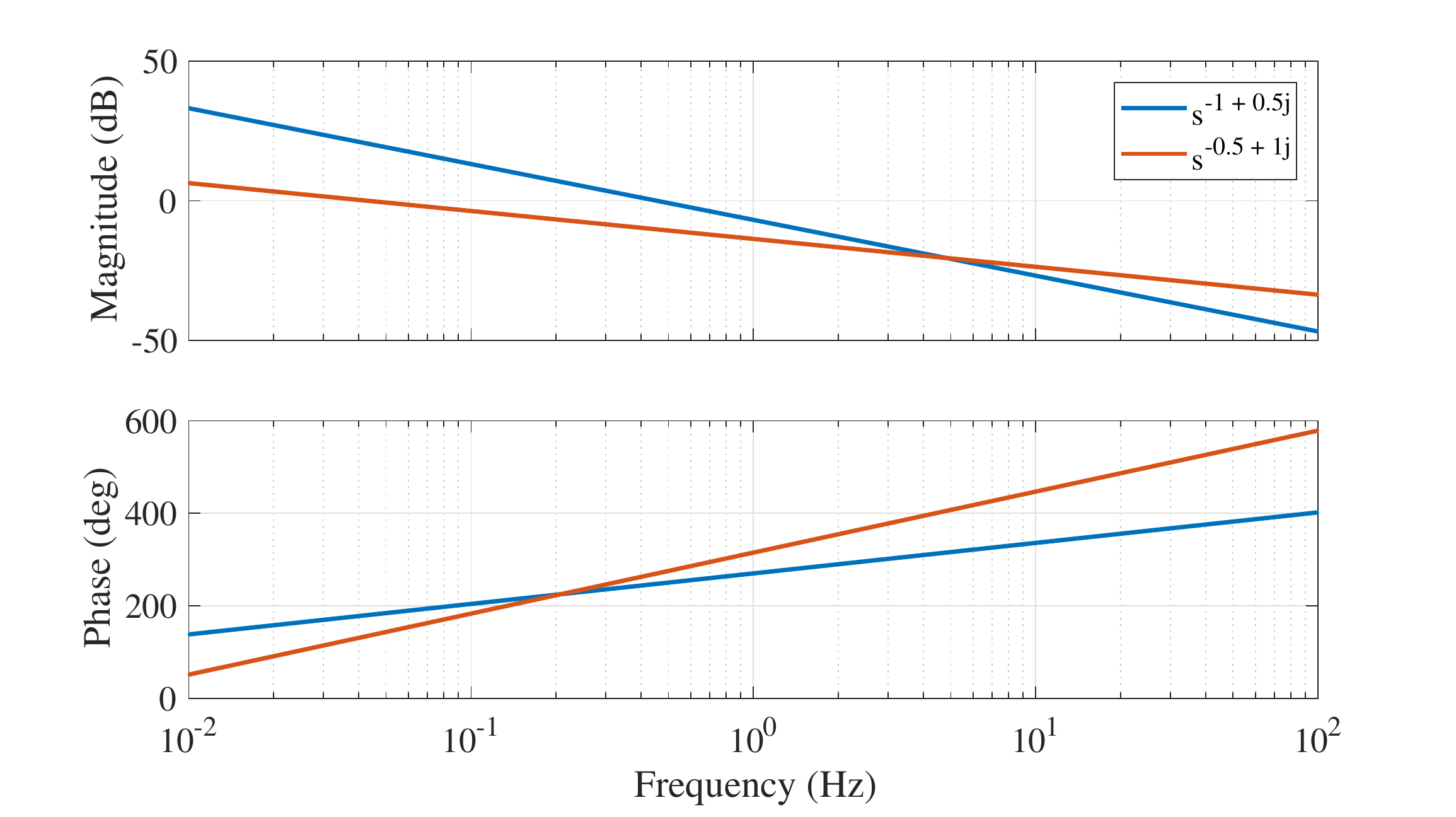}
	\caption{Frequency behaviour of complex order derivatives with order having negative real part and positive imaginary part}
	\label{fig:complexorder}
\end{figure}

When $\alpha < 0, \beta > 0$, the gain has a negative slope with a corresponding positive phase slope with examples shown in Fig. \ref{fig:complexorder}. As noted earlier, this behaviour is highly desirable to improve the open-loop shape. The CRONE approximation of this behaviour proposed in \cite{moghadam2018tuning} often leads to unstable poles or non-minimum phase zeros in the controller and thus in open-loop which are unacceptable from the point of view of industrial implementation. While CRONE logarithmic phase method \cite{oustaloup1999commande} allows only real and stable poles along with minimum phase zeros, the approximation is good only with respect to phase and a positive gain slope is seen resulting in no improvement compared to a linear lead filter. The disadvantages of proposed solutions in literature in approximating $s^{\alpha+j\beta}$ necessitate the work in this paper where nonlinear reset elements explained in the next subsection are used for this purpose.

\subsection{Reset control definition}
Reset controller is a type of impulsive system with a linear configuration at its base and an additional reset mechanism which is provoked when the resetting condition is satisfied. This reset action introduces nonlinearity and also the much-coveted phase advantage \cite{Banos2012}. Such a reset controller (including resetting and non-resetting parts) combined with a plant in series to obtain open-loop can be represented as
\begin{align*}
	\label{eq:reset}
	\dot{x}&=Ax+Be,&x,e,t\notin\mathcal{M}\\
	x(t^+)&=A_R x,&x,e,t\in\mathcal{M} \numberthis \\
	y&=Cx+De& 
\end{align*}
where $e$ is the input error signal, $y$ is plant output, $x=[x_R^T,x_p^T]^T$ is the state vector with $x_R$ and $x_p$ denoting reset controller and plant states respectively, $x_R(t)=[x_r^T\;x_{nr}^T]^T$ where $x_r \in \mathbb{R}^{n_r\times 1}$ and $x_{nr} \in \mathbb{R}^{n_{nr}\times 1}$ are states of the reset element ($C_r$) and non-resetting linear controller ($C_{nr}$) respectively. The continuous flow dynamics of the linear base are represented by the first and third equations with the second equation denoting the jump (reset) action which introduces the nonlinearity. 

The jump (reset) mechanism is controlled by resetting matrix $A_R$. $A_R=[{A_{\rho}}_{n_r\times n_r} 0;0\; I_{n_{nr}}]$. Note that $A_{\rho}$ is a diagonal matrix and need not be a zero matrix resulting in generalization of reset controllers in terms of control over after-reset value. Hence, $A_\rho = diag(\gamma_1, \gamma_2, ... \gamma_{n_r})$. In general $-1 \leq \gamma_i \leq 1$. Various reset conditions have been studied in literature with the most popular one being zero crossing of error input, i.e., $e(t) = 0$ owing to advantages in analysis and tuning, and hence this reset condition is also used in this paper. 

\subsection{Describing Function (DF)}
The main advantages of reset controller over other nonlinear options are that they are inherently integrable into the PID-framework and that their behaviour in frequency domain can be approximated using describing function enabling design using loop-shaping. DF is a quasi-linearisation technique with sinusoidal inputs considered in this case. The analytical equations to obtain DF for reset based systems with reset condition $e(t) = 0$ is provided in \cite{Guo2009} as
\begin{align}
\label{eq:DF}
G(j\omega)=C^T(j\omega I-A)^{-1}(I+j\Theta_D(\omega))B+D
\end{align}
where
\begin{align*}
\Theta_D(\omega)&\buildrel{\Delta}\over{=}-\frac{2\omega^2}{\pi}\Delta(\omega)(\Gamma_R(\omega)-\Lambda^{-1}(\omega)) \numberthis \\
\Lambda (\omega) &\buildrel{\Delta}\over{=}\omega^2+A^2\\
\Delta(\omega)&\buildrel{\Delta}\over{=}I+e^{\frac{\pi}{\omega}A}\\
\Delta_R(\omega)&\buildrel{\Delta}\over{=}I+A_Re^{\frac{\pi}{\omega}A}\\
\Gamma_R(\omega)&=\Delta_R^{-1}(\omega)A_R\Delta(\omega)\Lambda^{-1}(\omega)
\end{align*}
 
\begin{figure}
	\centering
	\includegraphics[trim = {1cm 0 1cm 0}, width=\linewidth]{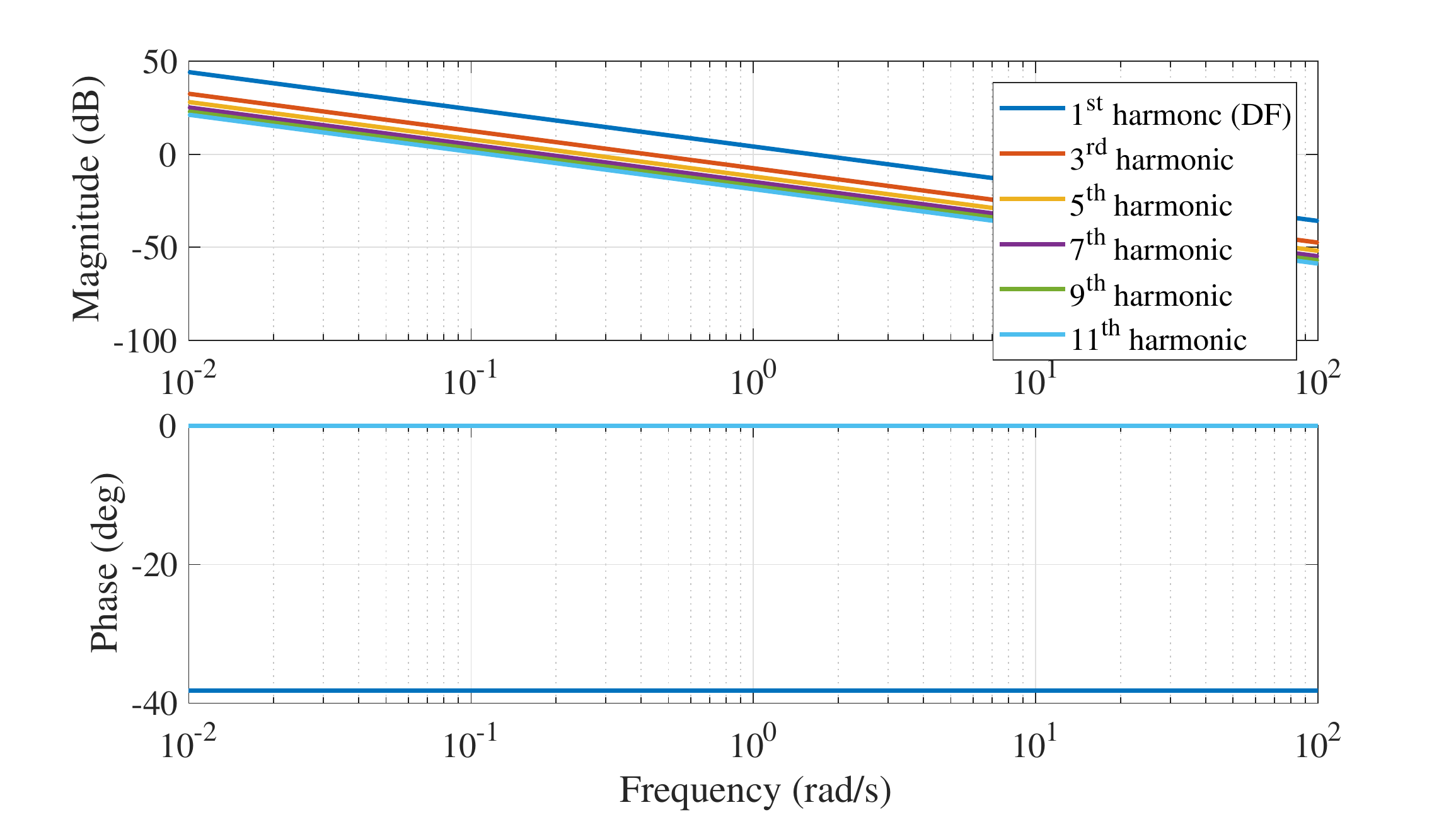}
	\caption{Frequency behaviour of Clegg integrator showing higher order harmonics up-to 11\textsuperscript{th} order. The even harmonics are zero and hence not shown.}
	\label{fig:clegghigh}
\end{figure}

Recently, the idea of higher-order sinusoidal input describing functions (HOSIDFs) for nonlinear systems introduced by \cite{NUIJ20061883} was extended to reset systems by \cite{Kars} to obtain analytical equations for HOSIDFs as
\begin{align*}
	G(j\omega,n)=&\begin{cases}
		\frac{-2\omega^2C}{j\pi}(A-j\omega nI)^{-1}\Delta(\omega)\big[\Gamma_R(\omega)-\Lambda^{-1}(\omega)\big]B, \numberthis \label{eq:HOSIDF}\\
		\mbox{\ \ \ \ \ \ \ \ \ \ \ \ \ \ \ \ \ \ \ \ \ \ \ \ \ \ \ \ \ \ \ \ \ for odd }n\geq 2\\
		0,\mbox{\ \ \ \ \ \ \ \ \ \ \ \ \ \ \ \ \ \ \ \ \ \ \ \ \ \ \ \ \ \ for even }n\geq 2
	\end{cases}
\end{align*}
where $n$ denotes the order of the harmonics. These equations are used to obtain the frequency behaviour of the much used Clegg integrator including the higher-order harmonics up-to the 11\textsuperscript{th} order in Fig. \ref{fig:clegghigh}. While it is seen that the magnitude of the higher-order harmonics is quite less compared to the first harmonic providing impetus to the idea of using DF for design and analysis of reset systems, these harmonics are not completely negligible as is done currently in literature and need to be taken into consideration during performance analysis.

Stability conditions for reset systems have been studied in literature extensively. We refer the readers to the works of \cite{beker2004fundamental} for $H_\beta$ conditions and the work of \cite{nevsic2008stability} using piece-wise Lyapunov equations in this regard. Since this is not the focus of the paper, this is not elaborated upon further for reasons of brevity.

\subsection{Constant-gain Lead-phase (CgLp)}

Recently, `Constant-gain Lead-phase' (CgLp) filter was presented to obtain constant unity gain with phase lead in the required frequency range, thereby enabling the use of reset for broadband phase compensation. CgLp integrated into PID-framework results in better open-loop shape; and significant improvements in tracking and steady-state precision are reported in \cite{arun2018,saikumar2018constant}. CgLp is designed by using a reset first (or second)-order lag filter $R$ in series with a corresponding linear first (or second)-order lead filter $L$ where the two filters are designed to have the same cut-off frequency as given below.
\begin{equation}
R(s) = \dfrac{1}{\cancelto{A_\rho}{(s/\omega_{r\alpha})^2 + (2s\beta_r/\omega_{r\alpha}) + 1}} \ \ \ \ \ \ \mbox{or}\  \dfrac{1}{\cancelto{A_\rho}{s/\omega_{r\alpha} + 1}}
\label{Reset_lag}
\end{equation}
and 
\begin{equation}
L(s) = \dfrac{(s/\omega_r)^2 + (2s\beta_r/\omega_r) + 1}{(s/\omega_f)^2 + (2s/\omega_f) + 1} \ \ \ \ \ \ \mbox{or}\ \dfrac{s/\omega_r + 1}{s/\omega_f + 1}
\end{equation}
where the arrow indicates the resetting action. Although both filters are to be designed to have same cut-off frequency, it can be seen that a different term $\omega_{r\alpha}$ is used for filter $R$. This is to account for the shift in the corner frequency of reset elements as noted in detail in \cite{saikumar2018constant}. An LPF at $\omega_f$ is primarily used to make filter $L$ proper.

\begin{figure}
	\centering
	\includegraphics[trim = {1cm 0 1cm 0}, width=1\linewidth]{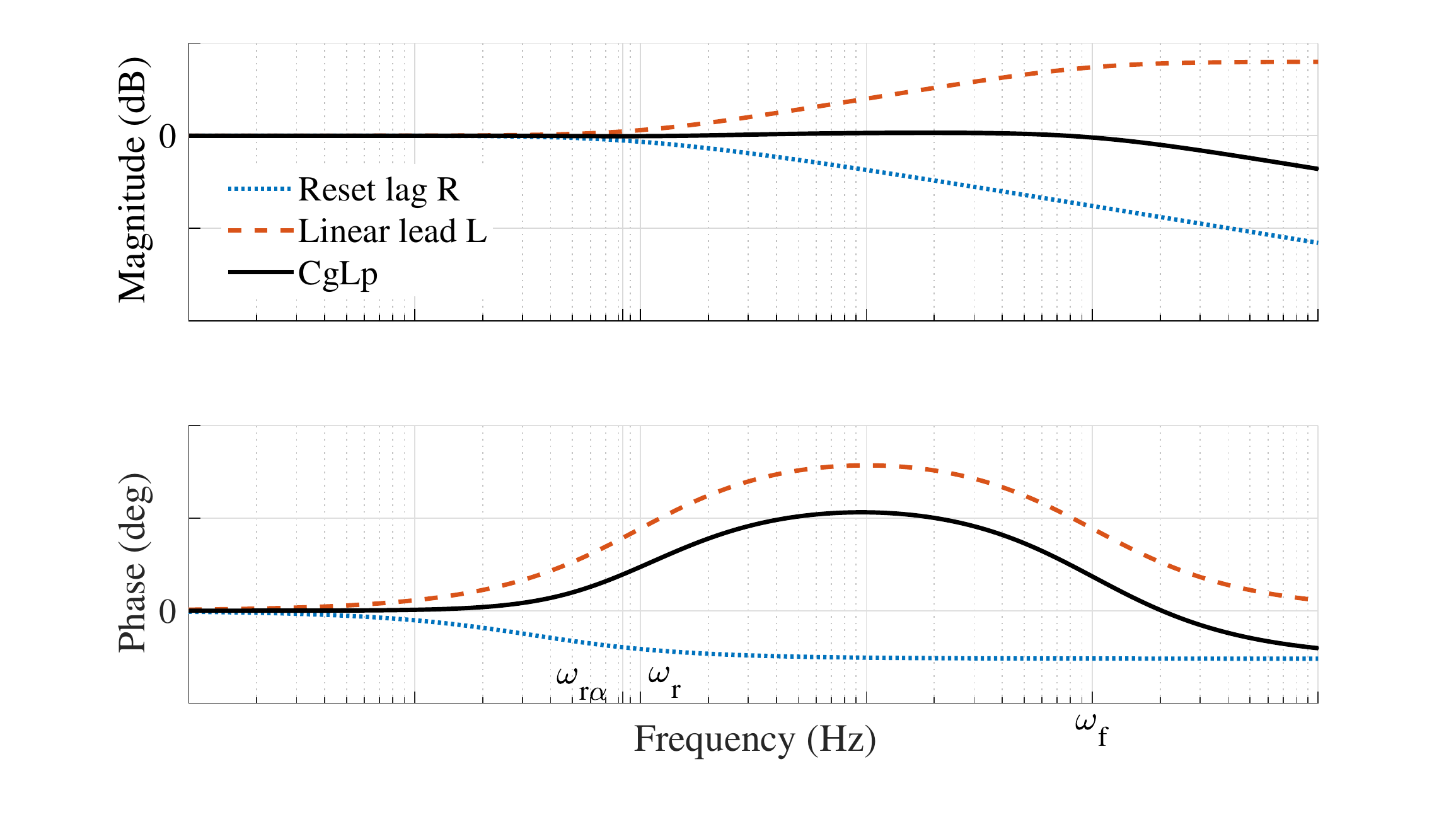}
	\caption{Phase lead achieved with constant gain with CgLp based on describing function analysis}
	\label{fig:CgLp}	
\end{figure}

A fundamental property of reset filters based on describing function is that the gain behaviour is not significantly affected while significant phase lag reduction is seen compared to the linear counterparts. As a result in CgLp, the gain behaviour of $R$ and $L$ cancel each-other resulting in unity gain, while the reduced phase lag of $R$ combined with the phase lead of $L$ results in phase lead as shown in Fig. \ref{fig:CgLp}.

%%%%%%%%%%%%%%%%%%%%%%%%%%%%%%%%%%%%%%%%%%%%%%%%%%%%%%%%%%%%%%%%%%%%%%%%%%%%%%%%
\section{COMPLEX ORDER CONTROLLER}
\label{cloc}

While CgLp filters integrated into PID result in performance improvement, CLOC, where the derivative action can be replaced by a complex order filter having a negative gain slope and a positive increasing phase should result in even further improved open-loop shape and performance. The design of a nonlinear filter whose describing function behaviour imitates the frequency response of a complex order derivative in a given frequency range is explained in this section.

Conceptually, the design of complex order filters follows from the design of CgLp. In CgLp, the reset lag and linear lead filters are placed at the same frequency to obtain cancellation in gain. This is true even in the case of second order filters. However, if the relative location of the poles and zeros are varied on the frequency axis, and with a corresponding introduction of reset for the poles, negative gain slope along with phase lead can be achieved according to describing function.

\subsection{Reset for complex order behaviour}

A real non-integer order derivative $s^\alpha$ can be approximated in the required frequency range $[\omega_l, \omega_h]$ of interest, with $N$ stable real poles and $N$ real minimum phase zeros as
\begin{align}
s^\alpha &\approx C \prod_{m=1}^{N} \frac{1+\frac{s}{\omega_{z,m}}}{1+\frac{s}{\omega_{p,m}}}
\label{eq:crone} \\
\omega_{z,m} &= \omega_l \left( \frac{\omega_h}{\omega_l} \right)^{\frac{2m-1-\alpha}{2N}}
, 
\omega_{p,m} &= \omega_l \left( \frac{\omega_h}{\omega_l} \right)^{\frac{2m-1+\alpha}{2N}}
\label{eq:crone:poleszeros}
\end{align}

The poles and zeros are placed in pairs such that with an appropriate value of $N$, the asymptotic behaviour is smoothed out to obtain $20\alpha\mbox{ dB/decade}$ gain slope and an almost constant $\frac{\alpha\pi}{2}\mbox{rad/s}$ phase in the range $[\omega_l, \omega_h]$. This approximation works well for real non-integer orders and a negative value of $\alpha$ results in a negative gain slope, but also a negative phase. In order to obtain the required complex order filter behaviour with order $\alpha + j\beta$, reset mechanism is introduced into this approximation as follows.

For a given complex order $\alpha + j\beta$ to be approximated in frequency range $[\omega_l,\omega_h]$:
\begin{enumerate}
	\item Obtain the values of $\omega_{p,m}$ and $\omega_{z,m}$ using Eqns. \ref{eq:crone:poleszeros}.
	\item The transfer function of Eqn. \ref{eq:crone} which is used to approximate $s^\alpha$ is split into resetting and non-resetting parts as:
	\begin{align}
	C_r = \prod_{m=1}^{N} \frac{1}{1+\frac{s}{\omega_{p,m}}}\\
	C_{nr} = \prod_{m=1}^{N} \frac{1+\frac{s}{\omega_{z,m}}}{1+\frac{s}{\tilde{\omega}_{p,m}}}
	\end{align}
	where $\tilde{\omega}_{p,m}$ poles are used to make the non-resetting part proper. These values can be chosen to be well above $\omega_h$ to ensure effectively zero influence. \\
	\item Choose an appropriate resetting matrix $A_\rho$ for $C_r$ to ensure that the combined frequency response of $C_r$ and $C_{nr}$ obtained through describing function approximates both required negative gain slope and positive phase slope.
\end{enumerate}

This results in 
\begin{align}
s^{\alpha+j\beta} \approx C \left( \cancelto{A_\rho}{ \prod_{m=1}^{N} \frac1{1+\frac{s}{\omega_{p,m}}} } \quad \quad \right)
\prod_{m=1}^{N} \frac{1+\frac{s}{\omega_{z,m}}}{1+\frac{s}{\tilde{\omega}_{p,m}}}
\label{eq:reset_alpha}
\end{align}

\begin{figure}
	\centering
	\includegraphics[trim = {1cm 0 1cm 0}, width=1\linewidth]{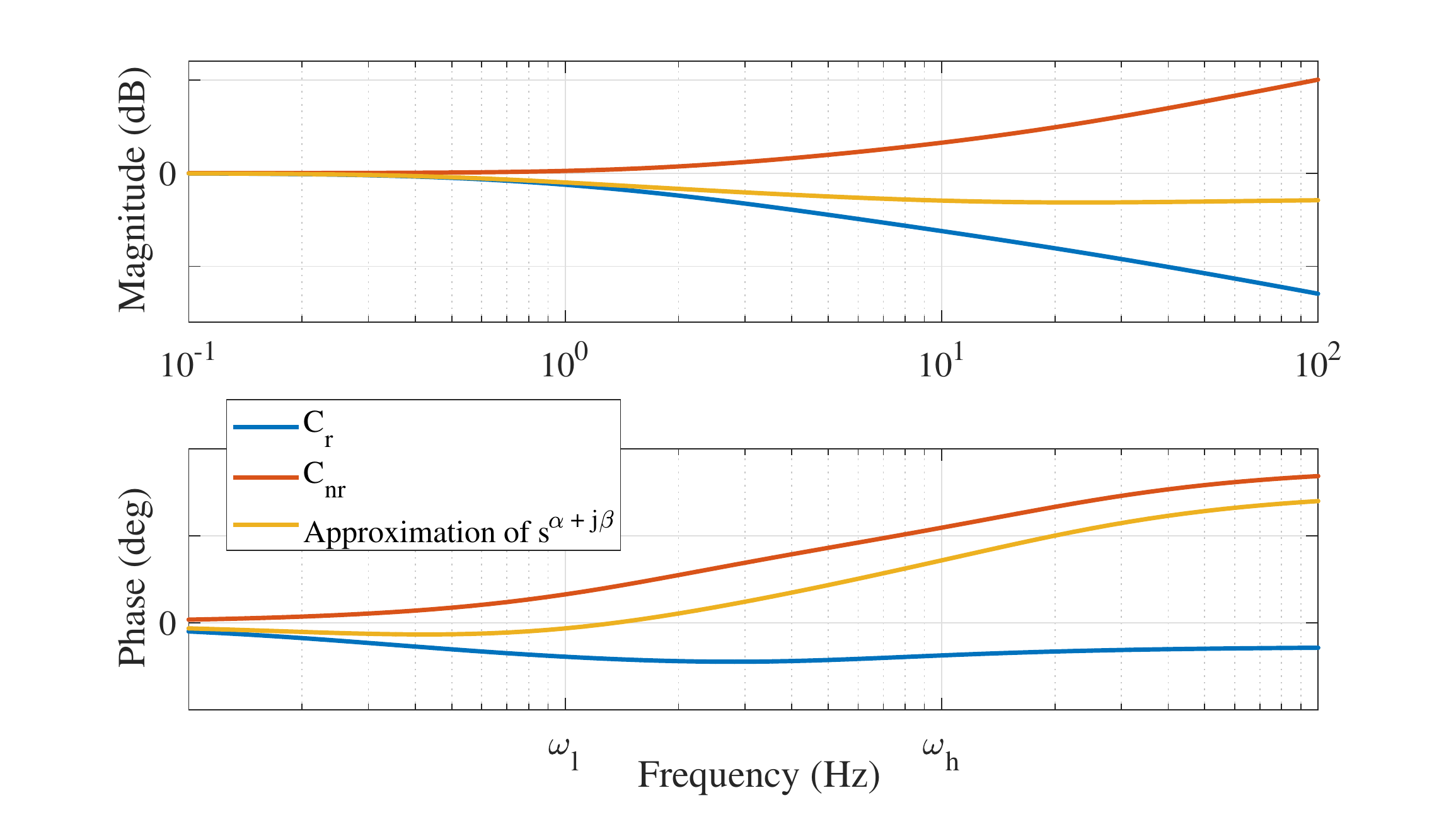}
	\caption{Positive increasing phase achieved with negative gain slope based on describing function analysis. $N = 2$ is used in this case with a heuristically chosen $\alpha$ and $A_\rho$ matrix.}
	\label{fig:cloc}	
\end{figure}

Similar to the case of CgLp, the introduction of reset only for the poles of the approximated transfer function does not significantly change the gain behaviour. However, it has a significant influence on the phase lag of each pole, which is certainly less than that of its linear counterpart. While in the case of CgLp, a single lead filter introduces phase lead and combined with the reduced phase lag of the reset lag filter, produces an overall phase lead; in the case of the presented approximation, each of the zeros of $C_{nr}$ provides phase lead which when combined with the reduced phase lag of each of the reset poles results not only in phase lead but also positive slope for phase. This is shown using an example in Fig. \ref{fig:cloc}.

\subsection{Tuning of $A_\rho$}

While the general principle behind approximating frequency behaviour of a complex order derivative is explained in the previous subsection, the calculation of $A_\rho$ matrix to achieve the required phase slope is not straight-forward. The complexity arises from the resetting of multiple poles. Since reset filters are nonlinear, the effect of introducing reset to two filters cannot be estimated in a linear fashion. This can be easily seen in the case of FORE which has a phase lag of $38.1^\circ$, while a SORE has a phase lag of $51.8^\circ$. This problem is further exacerbated when the diagonal elements of $A_\rho$, i.e., $\gamma_i$, which correspond to the reset factor of each individual pole are not equal.

As such, we use a numerical approach to obtain the values of $\gamma_i$ in the $A_\rho$ matrix to achieve the required phase slope as follows.
\begin{enumerate}
	\item For every combination of $\gamma_i \in -1:\delta:1$, obtain describing function of $C_r$.
	\item $C_{nr}$ remains unchanged and the frequency response of this linear filter is combined with the describing function of $C_r$ to obtain the describing function based frequency response of the reset based approximation.
	\item Choose the combination of $diag(\gamma_1, \gamma_2... \gamma_N)$ that best approximates both gain slope and phase slope.
\end{enumerate}

The reason for comparing both obtained gain and phase slope against required values is because the resetting action results in a small change in the pole location according to describing function as explained in \cite{saikumar2018constant}. Although this change is not significant in most cases especially when values of $\gamma_i$ are not close to -1, the use of multiple reset filters could potentially modify the gain slope sufficiently to cause errors. To further accurately obtain the values of $A_\rho$ which best approximates the required behaviour, linear interpolation between best combinations could be used.

%%%%%%%%%%%%%%%%%%%%%%%%%%%%%%%%%%%%%%%%%%%%%%%%%%%%%%%%%%%%%%%%%%%%%%%%%%%%%%%%
\section{EXPERIMENTAL SETUP}
\label{experimental}

\begin{figure}
	\centering
	\includegraphics[width= 0.78\linewidth]{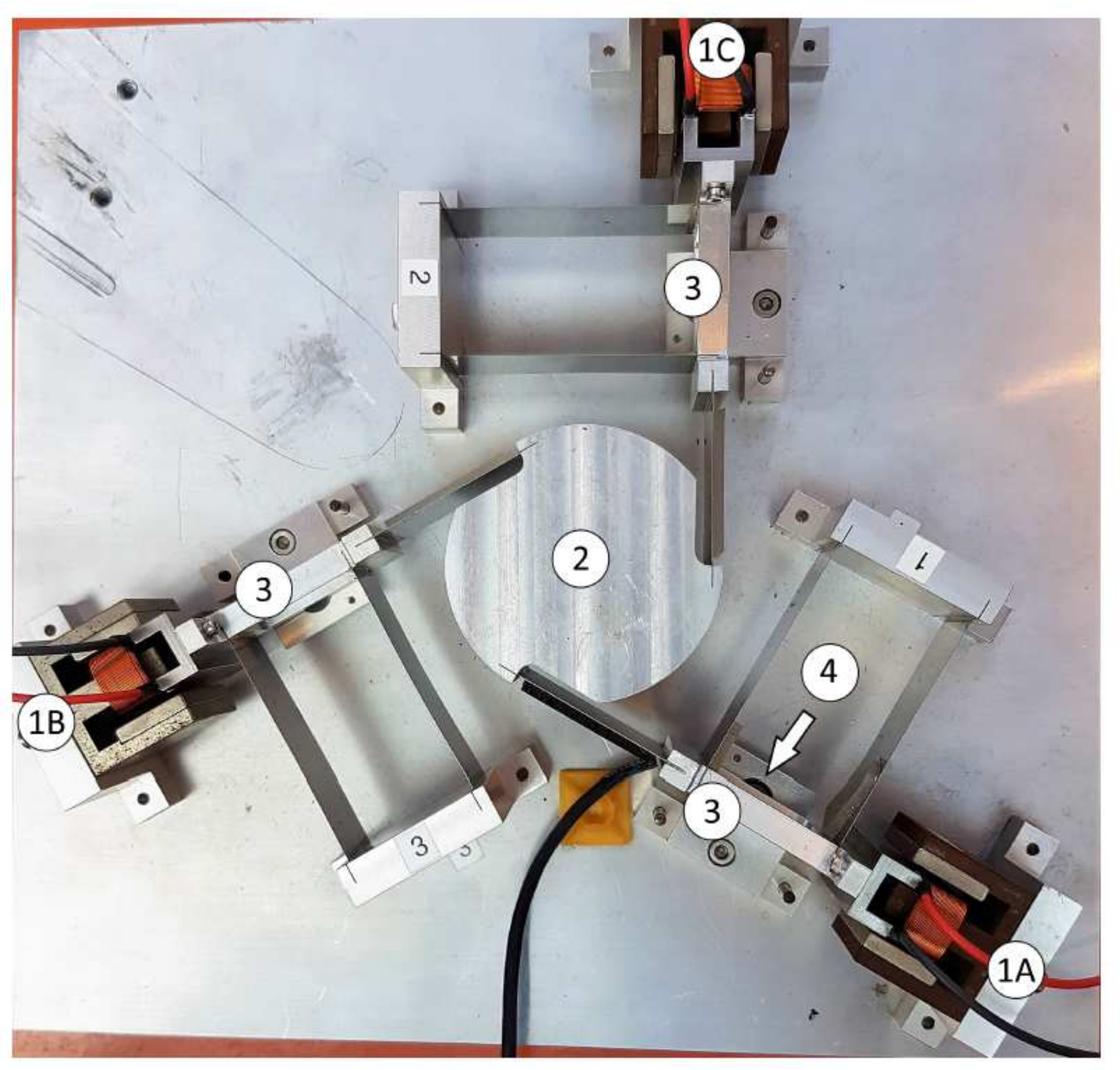}
	\caption{Planar precision positioning `Spyder' stage with voice coil actuators denoted as 1A, 1B and 1C controlling the three masses (indicated as 3) and constrained by leaf flexures. The central mass (indicated by 2) is connected to these 3 masses through leaf flexures and linear encoders (indicated by 4) placed under masses `3' provide position feedback with resolution of $100\mbox{ nm}$.}
	\label{fig:setup}	
\end{figure}

\begin{figure}
	\centering
	\includegraphics[trim = {1cm 0 1cm 0}, width=\linewidth]{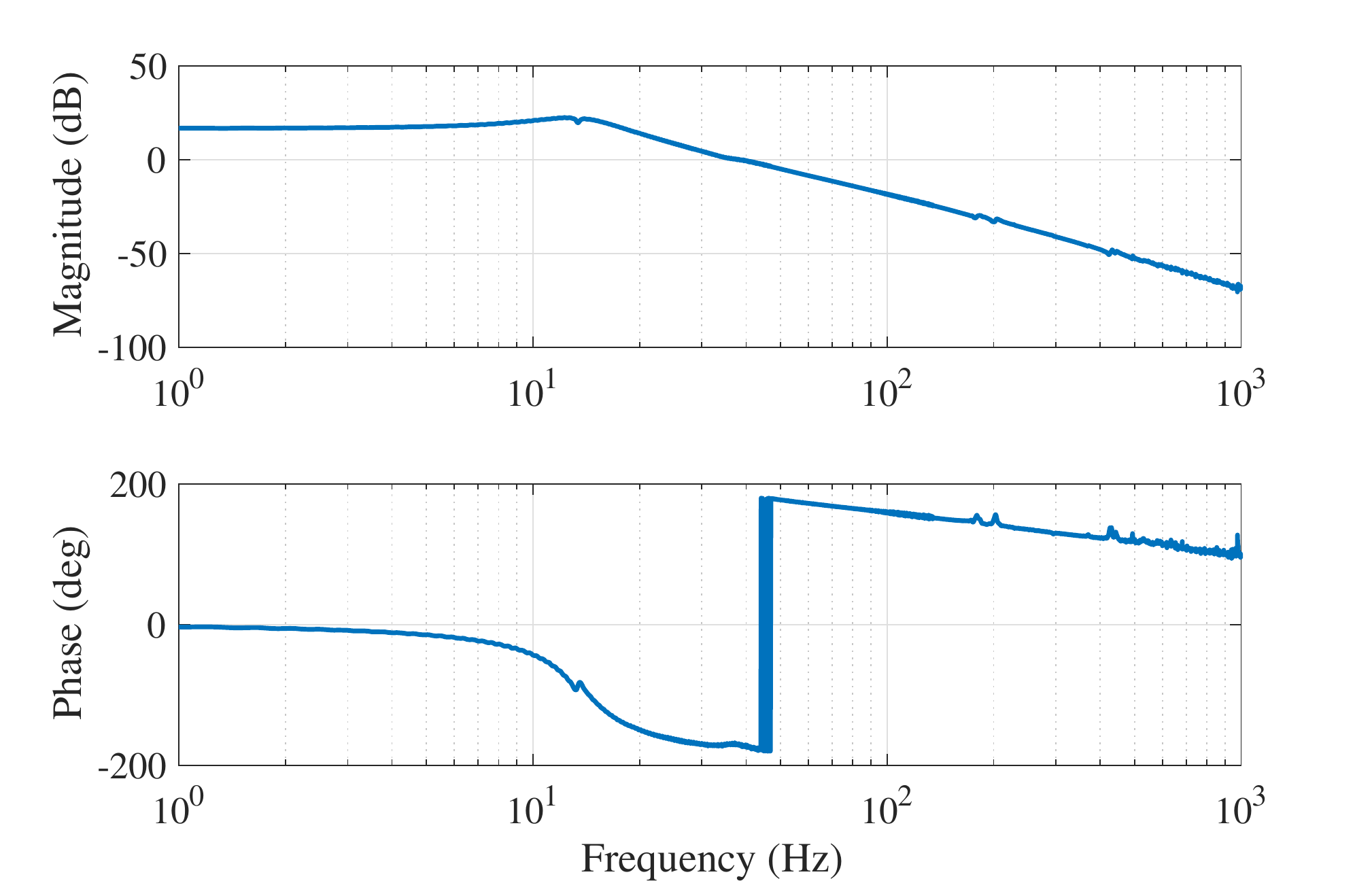}
	\caption{Frequency response data of plant as seen from actuator '1A' to position of mass '3' attached to same actuator.}
	\label{fig:frf}	
\end{figure}

The planar precision positioning stage named `Spyder' shown in Fig. \ref{fig:setup} is used for validation of the designed controllers and performance analysis. Since we are dealing with the design of SISO controllers, only one of the actuators (1A) is used to control the position of mass `3' rigidly attached to the same actuator. All the controllers are implemented using FPGA on a NI CompactRIO system to achieve real-time control with a sampling frequency of $10\mbox{ kHz}$. LM388 linear power amplifiers are used as the drive with a Mercury M2000 linear encoder providing position sensing with a resolution of $100\mbox{ nm}$. The frequency response function (frf) of the plant is obtained (shown in Fig. \ref{fig:frf}) to be in line with the industry technique of using the frf directly to shape the open-loop. The plant's frequency behaviour is similar to that of a collocated double mass-spring-damper with additional dynamics at frequencies much higher that of the first eigenmode. This can be simplified to a second-order plant as given in Eqn. \ref{eq:system} for use in stability analysis. Furthermore, this estimated transfer function is also used to create a simple feed-forward by inverting the transfer function and making it strictly proper. All feedback controllers are designed to ensure a cross-over frequency of $150\mbox{ Hz}$ and a phase margin of $30^\circ$.

\begin{equation}
\label{eq:system}
G(s) = \frac{1.429\times10^8}{175.9s^2 + 7738s + 1.361\times10^6}
\end{equation}

In keeping with the idea of improving the open-loop shape with nonlinear filters, five different controllers are designed with progressive improvement seen in the open-loop frequency response. As benchmark, a linear PID controller is designed with transfer function provided as:
\begin{equation}
\label{eqPID}
PID = K_p\Bigg(\frac{s + \omega_i}{s}\Bigg)\Bigg(\frac{1 + \frac{s}{\omega_d}}{1 + \frac{s}{\omega_t}}\Bigg)\Bigg(\frac{1}{1 + \frac{s}{\omega_f}}\Bigg)
\end{equation}
where $\omega_i$ is the frequency at which integrator action terminates, $\omega_d$ and $\omega_t$ are the starting and  taming frequencies of differentiator (lead filter), and $\omega_f$ is corner frequency of low pass filter used to attenuate noise at high frequencies with $\omega_i < \omega_d < \omega_t < \omega_f$. In general $\omega_d$ and $\omega_t$ are chosen as $\omega_d = \omega_c/a$ and $\omega_t = a\omega_c$ where $a > 1$ to ensure that the peak of phase lead achieved with the differentiator action coincides with the chosen open-loop cross-over frequency $\omega_c$. $K_p$ is adjusted to ensure the said cross-over frequency of $\omega_c$.

Since CgLp with constant gain and phase lead provides an improvement in open-loop shape, two CgLp based controllers (one using FORE and another with SORE) are designed as an intermediate step between PID and complex order controller. Since FORE based CgLp is not capable of providing the required phase lead, a linear lead filter is also used to make up for the remaining phase and this controller is referred to as CgLp-PID. However, SORE based CgLp is capable of providing more phase lead than required and hence is tuned to ensure that required PM is achieved and the lead filter is completely unused resulting in $a = 1$ in the calculation of $\omega_d$ and $\omega_t$. This controller is referred to as CgLp-PI since the linear derivative filter is not used.

\begin{figure}
	\centering
	\includegraphics[trim = {1cm 0 1cm 0}, width=1\linewidth]{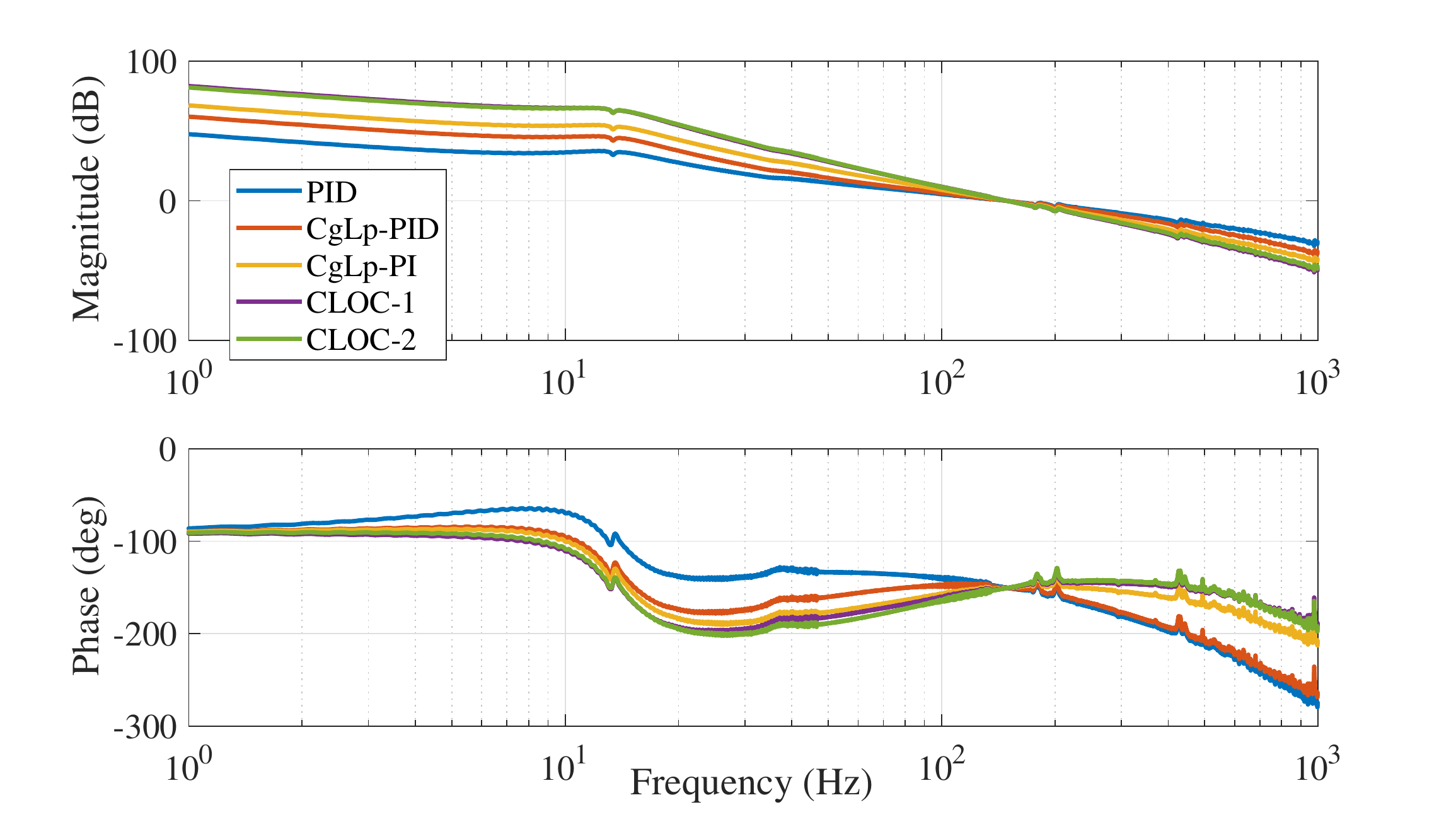}
	\caption{Open-loop frequency response with all 5 designed controllers plotted using frf of plant and describing function for controllers. This shows the expected improvement in tracking and steady-state precision with improved loop shape.}
	\label{fig:1harmonic}	
\end{figure}

Finally, two different complex order derivatives are approximated with $N = 3$ and combined with PI controllers to obtain CLOC-1 and CLOC-2. The complex order derivative in CLOC-1 is designed to approximate a complex order derivative with order $-0.5 + 0.9475j$ resulting in a gain slope of $-10\mbox{ dB/decade}$ and phase slope of $125^\circ\mbox{/decade}$. This derivative is approximated over two decades centred around $\omega_c$. Similarly, complex order derivative of CLOC-2 has order $-0.5 + 1.1370j$ which is approximated over one and a half decades to obtain a gain slope of $-10\mbox{ dB/decade}$ and a phase slope of $150^\circ\mbox{/decade}$. These values are heuristically chosen and the obtained CLOC controllers are combined with the plant to check for stability before implementation. The complete details of all the controllers are provided in Table. \ref{Tab:controller}. The open-loop frequency response obtained using frf of the plant and the describing function behaviour of the controllers is shown in Fig. \ref{fig:1harmonic}. Assuming DF based analysis is accurate, PID has the worst open-loop shape while the shapes of both CLOC-1 and CLOC-2 are close to each other and should provide the best performance in tracking and steady-state precision. 

\begin{table}
	\centering
	\begin{tabular}{|c||l|}
		\hline
		PID & $\omega_d = 16.43\mbox{ Hz}$, $\omega_t = 1369.3\mbox{ Hz}$ \\ \hline
		CgLp-PID  & $\omega_r = 50\mbox{ Hz}$, $\omega_{r\alpha} = 35.7\mbox{ Hz}$ \\  
		 & $\omega_d = 68.4\mbox{ Hz}$, $\omega_t = 328.8\mbox{ Hz}$ \\ \hline
		CgLp-PI & $\omega_r = 78.9\mbox{ Hz}$, $\omega_{r\alpha} = 68.6138\mbox{ Hz}$ \\ \hline
		CLOC-1 & $\omega_{p,m} = [16.5, 76.6, 355.5]\mbox{ Hz}$ \\
		& $\omega_{z,m} = [35.55, 165.0, 766.0]\mbox{ Hz}$ \\
		& $A_\rho = diag(0.21, -0.22, 0.1)$ \\ \hline
		CLOC-2 & $\omega_{p,m} = [27.0, 85.4, 270.0]\mbox{ Hz}$ \\
		& $\omega_{z,m} = [48.0, 151.8, 480.3]\mbox{ Hz}$ \\
		& $A_\rho = diag(0.29, -0.26, 0.3)$ \\ \hline
	\end{tabular}
	\caption{Relevant parameters of all designed controllers. $\omega_i = 15\mbox{ Hz}$ is used in all cases. $K_p$ is appropriately adjusted to ensure cross-over occurs at $150\mbox{ Hz}$. All values are given in $\mbox{Hz}$ and are converted to $\mbox{rad/s}$ in the equations for calculation and implementation.}
	\label{Tab:controller}
\end{table}

%%%%%%%%%%%%%%%%%%%%%%%%%%%%%%%%%%%%%%%%%%%%%%%%%%%%%%%%%%%%%%%%%%%%%%%%%%%%%%%%
\section{RESULTS AND DISCUSSION}
\label{results}

\begin{figure}
	\centering
	\includegraphics[trim = {1cm 0 1cm 0}, width=1\linewidth]{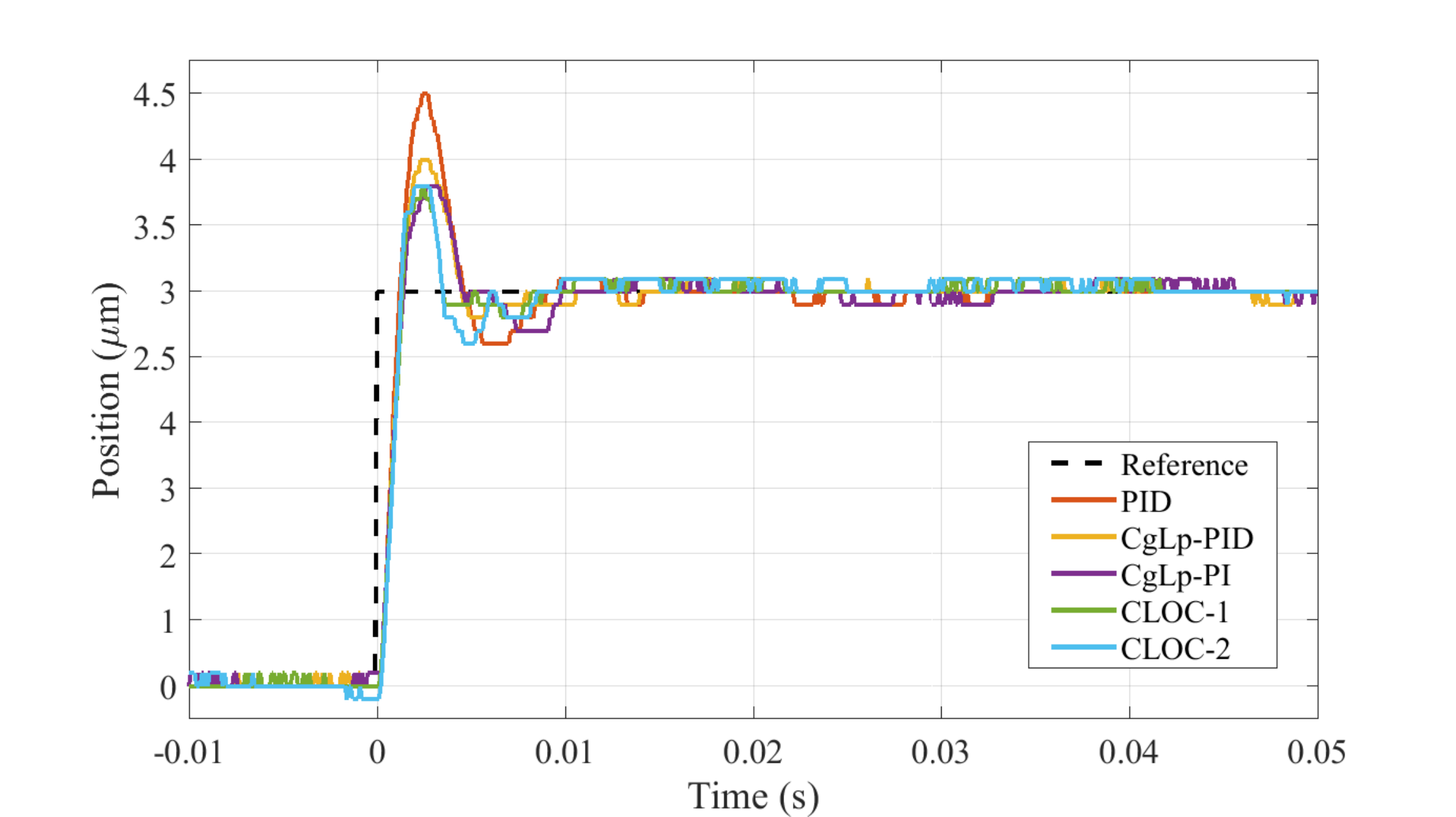}
	\caption{Step response of all 5 controllers obtained for a step reference change of $3 \mu m$}
	\label{fig:steps}	
\end{figure}

The performance of all controllers is tested on the stage in both tracking and steady-state precision. The step responses of the controllers are also obtained and plotted in Fig. \ref{fig:steps}. In the case of steady-state precision measurements, it was noticed that the precision of all 5 controllers was within the resolution of the sensor. Hence, to obtain an idea of noise attenuation, additional white noise of amplitude $2\ \mu\mbox{m}$ was introduced into the sensor signal and the corresponding effect on precision was recorded; with the results tabulated in Table. \ref{Tab:performance}.

In the case of reference tracking, 3 different fourth-order pre-filtered trajectories designed as explained in \cite{lambrechts2005trajectory} are used for testing. All three cases result in a scanning motion of $100 \mu\mbox{m}$ range as is typical in several precision positioning stages, with the difference in the references being the rate of change of reference. While reference-1 achieves one scan from position 0 to position $100 \mu\mbox{m}$ in $397\mbox{ ms}$, reference-2 and reference-3 achieve the same in $235\mbox{ ms}$ and $93\mbox{ ms}$ respectively, providing us with a good idea of performance improvements over varying speeds of operation. The RMS error and maximum error are calculated for all cases for all controllers and have been provided in Table. \ref{Tab:performance}.

\begin{table*}
	\centering
	\begin{tabular}{||c||c|c||c|c||c|c||c||}
		\hline
		Controller & $e_{RMS}$ & $max(|e|)$ & $e_{RMS}$ & $max(|e|)$ & $e_{RMS}$ & $max(|e|)$ & $max(|e|)$ \\ \hline
		& \multicolumn{7}{|c||}{($100 \mbox{ nm}$)} \\ \hline \hline
		& \multicolumn{2}{|c||}{Reference 1} & \multicolumn{2}{|c||}{Reference 2} & \multicolumn{2}{|c||}{Reference 3} & Steady-state \\ \hline \hline
		PID & 154.97 & 500 & 191.18 & 700 & 429.37 & 1100 & 1400 \\ \hline
		CgLp-PID (GFORE) & 144.23 & 500 & 188.83 & 600 & 344.29 & 1100 & 800 \\ \hline
		CgLp-PI (GSORE) & 133.53 & 500 & 170.42 & 600 & 297.05 & 900 & 1400 \\ \hline
		CLOC-PI 1 & 133.02 & 500 & 168.23 & 600 & 314.30 & 1000 & 1400 \\ \hline
		CLOC-PI 2 & 125.03 & 500 & 149.11 & 500 & 266.68 & 1000 & 1300 \\ \hline
	\end{tabular}
	\caption{Performance indices of all designed controllers.}
	\label{Tab:performance}
\end{table*}

The step responses in Fig. \ref{fig:steps} shows that while all 5 controllers are designed to have the same PM according to describing function analysis, different levels of overshoot is seen with reset based CgLp-PID, CgLp-PI and CLOC controllers showing lower overshoot. It is also interesting that in the case of CgLp-PID where the phase lead required from the controller is not completely provided by the reset filter, but additional phase is added using a linear lead, the overshoot is slightly higher than that seen with CgLp-PI and CLOC controllers where the complete phase lead is provided by the reset filters.

An analysis of the tracking (RMS error) and steady-state (Max. error) precision results of Table. \ref{Tab:performance} shows that the performance of all nonlinear controllers is better than or equal to that of PID. Excluding CLOC-1, the improvement in performance is progressive for tracking results as expected. However, CLOC-1 does not follow this pattern and improvement is also not seen in the case of steady-state precision. The describing function based open-loop shape of Fig. \ref{fig:1harmonic} indicates towards a progressive improvement from PID to CLOC. Further, the open-loop of CLOC-1 and CLOC-2 is similar and the disparity in the performance cannot be explained.

\begin{figure}
	\centering
	\includegraphics[trim = {1cm 0 1cm 0}, width=1\linewidth]{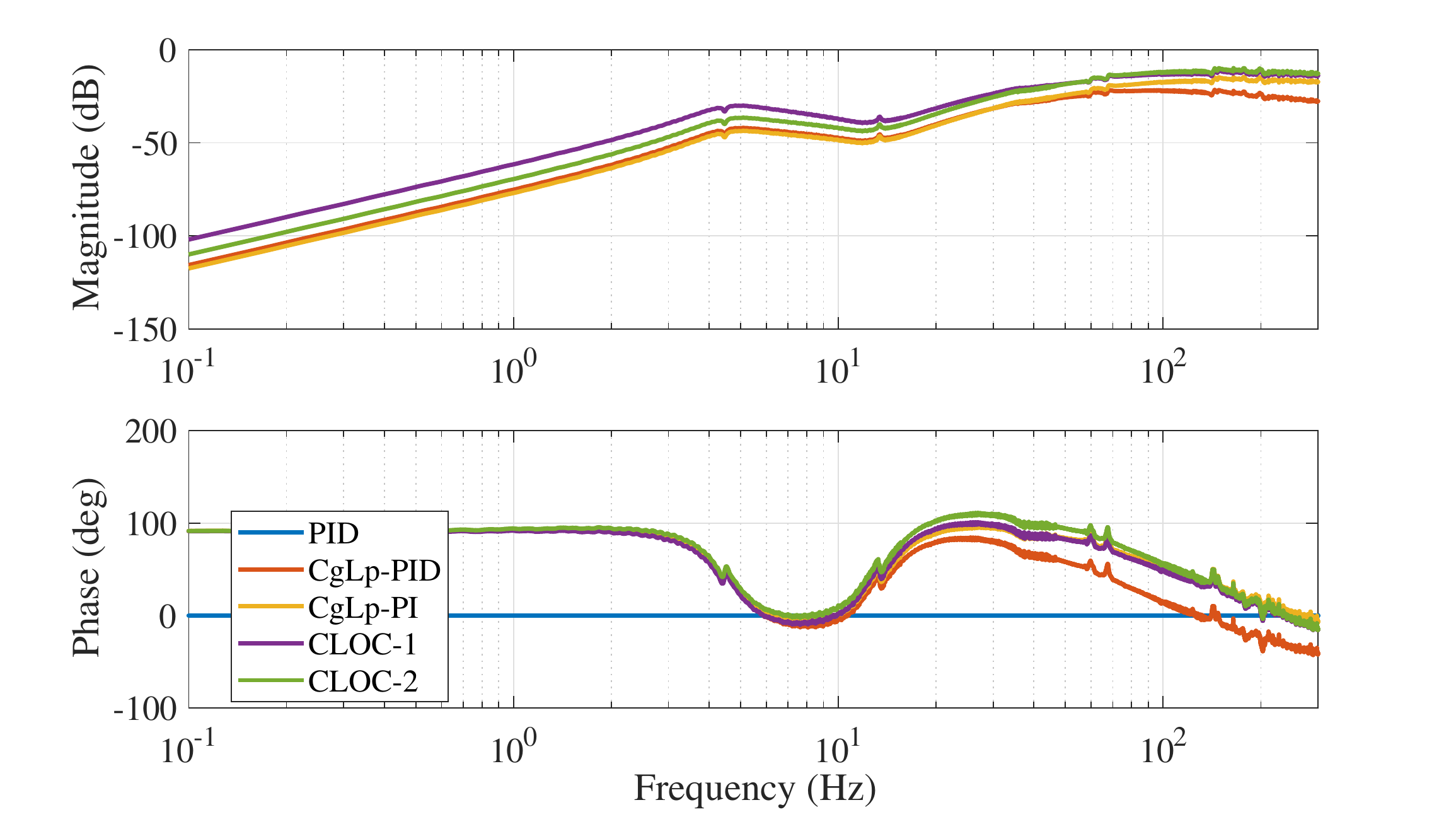}
	\caption{Open-loop third harmonic obtained using HOSIDF and normalized with respect to the first harmonic. This plot gives an idea of the relative performance deterioration to be expected from the higher-order harmonics.}
	\label{fig:3harmonicnormalized}	
\end{figure}

For better understanding, it is necessary to also study the higher order harmonics using the HOSIDF tool. Since the magnitude of the harmonics decreases with order as seen in Fig. \ref{fig:clegghigh}, we can study the third harmonic in open-loop. However, the open-loop of all controllers differ in both first and third harmonic with the former improving performance and the latter deteriorating performance.  Hence, the third harmonic is normalized with respect to the first harmonic and plotted in Fig. \ref{fig:3harmonicnormalized}. From this, it can be seen that the normalized third harmonic of CLOC-1 is significantly higher compared to the CgLp based controllers explaining the performance deterioration, while this is not the case for CLOC-2. Additionally, at higher frequencies, the magnitude is lower in the case of CgLp-PID compared to all the other nonlinear controllers, explaining the improvement seen in steady-state precision with the former and not with the latter.

%%%%%%%%%%%%%%%%%%%%%%%%%%%%%%%%%%%%%%%%%%%%%%%%%%%%%%%%%%%%%%%%%%%%%%%%%%%%%%%%
\section{CONCLUSION}
\label{conclusions}

The paper presents a practically implementable and applicable approximation of complex order derivative for the design of complex order controllers (CLOC) to provide improved loop-shaping. Reset control is used to introduce nonlinearity essential for the realization of this behaviour and the approximation based on describing function analysis. A numerical method to obtain the approximation in the required frequency range is presented. Different nonlinear controllers including two separate CLOC controllers which progressively show improvement in open-loop shape compared to the benchmark PID are designed for a precision planar positioning stage. While improvements in tracking and steady-state precision are seen, the results do not completely correlate with the open-loop shape. This disparity can be explained through the analysis of the third harmonic.

For future work, while the complex order derivatives are currently designed using a numerical approach, an analytical approach will aid speed up the design process especially if the value of $N$ used in approximation increases. Further, in terms of performance analysis, while the third harmonic plot provides an idea of comparative performance deterioration, this cannot be quantified. Hence, a relation between the open-loop of all harmonics and expected error in closed-loop needs to be developed to allow for better tuning of these controllers allowing for the stringent demands of the precision industry to be met.

\bibliographystyle{IEEEtran}
\bibliography{references}

\begin{thebibliography}{10}
\providecommand{\url}[1]{#1}
\csname url@samestyle\endcsname
\providecommand{\newblock}{\relax}
\providecommand{\bibinfo}[2]{#2}
\providecommand{\BIBentrySTDinterwordspacing}{\spaceskip=0pt\relax}
\providecommand{\BIBentryALTinterwordstretchfactor}{4}
\providecommand{\BIBentryALTinterwordspacing}{\spaceskip=\fontdimen2\font plus
\BIBentryALTinterwordstretchfactor\fontdimen3\font minus
  \fontdimen4\font\relax}
\providecommand{\BIBforeignlanguage}[2]{{%
\expandafter\ifx\csname l@#1\endcsname\relax
\typeout{** WARNING: IEEEtran.bst: No hyphenation pattern has been}%
\typeout{** loaded for the language `#1'. Using the pattern for}%
\typeout{** the default language instead.}%
\else
\language=\csname l@#1\endcsname
\fi
#2}}
\providecommand{\BIBdecl}{\relax}
\BIBdecl

\bibitem{aastrom2000limitations}
K.~J. {\AA}str{\"o}m, ``Limitations on control system performance,''
  \emph{European Journal of Control}, vol.~6, no.~1, pp. 2--20, 2000.

\bibitem{freudenberg2000survey}
J.~Freudenberg, R.~Middleton, and A.~Stefanpoulou, ``A survey of inherent
  design limitations,'' in \emph{Proceedings of the 2000 American Control
  Conference. ACC (IEEE Cat. No. 00CH36334)}, vol.~5.\hskip 1em plus 0.5em
  minus 0.4em\relax IEEE, 2000, pp. 2987--3001.

\bibitem{MSDbook}
R.~M. Schmidt, G.~Schitter, and A.~Rankers, \emph{The Design of High
  Performance Mechatronics-: High-Tech Functionality by Multidisciplinary
  System Integration}.\hskip 1em plus 0.5em minus 0.4em\relax Ios Press, 2014.

\bibitem{oustaloup2000frequency}
A.~Oustaloup, F.~Levron, B.~Mathieu, and F.~M. Nanot, ``Frequency-band complex
  noninteger differentiator: characterization and synthesis,'' \emph{IEEE
  Transactions on Circuits and Systems I: Fundamental Theory and Applications},
  vol.~47, no.~1, pp. 25--39, 2000.

\bibitem{sabatier2015fractional}
J.~Sabatier, P.~Lanusse, P.~Melchior, and A.~Oustaloup, ``Fractional order
  differentiation and robust control design,'' in \emph{CRONE, H-infinity and
  Motion Control}.\hskip 1em plus 0.5em minus 0.4em\relax Springer, 2015.

\bibitem{oustaloup1999commande}
A.~Oustaloup and B.~Mathieu, \emph{La commande {CRONE}}.\hskip 1em plus 0.5em
  minus 0.4em\relax HERMES science publ. Paris, 1999.

\bibitem{moghadam2018tuning}
M.~G. Moghadam, F.~Padula, and L.~Ntogramatzidis, ``Tuning and performance
  assessment of complex fractional-order pi controllers,''
  \emph{IFAC-PapersOnLine}, vol.~51, no.~4, pp. 757--762, 2018.

\bibitem{Clegg1958}
J.~Clegg, ``A nonlinear integrator for servomechanisms,'' \emph{Transactions of
  the American Institute of Electrical Engineers, Part II: Applications and
  Industry}, vol.~77, no.~1, pp. 41--42, 1958.

\bibitem{chait2002horowitz}
Y.~Chait and C.~Hollot, ``On {H}orowitz's contributions to reset control,''
  \emph{International Journal of Robust and Nonlinear Control: IFAC-Affiliated
  Journal}, vol.~12, no.~4, pp. 335--355, 2002.

\bibitem{zaccarian2005first}
L.~Zaccarian, D.~Nesic, and A.~R. Teel, ``First order reset elements and the
  clegg integrator revisited,'' in \emph{Proceedings of the 2005, American
  Control Conference, 2005.}\hskip 1em plus 0.5em minus 0.4em\relax IEEE, 2005,
  pp. 563--568.

\bibitem{Hazeleger2016}
L.~Hazeleger, M.~Heertjes, and H.~Nijmeijer, ``Second-order reset elements for
  stage control design,'' in \emph{American Control Conference (ACC),
  2016}.\hskip 1em plus 0.5em minus 0.4em\relax IEEE, 2016, pp. 2643--2648.

\bibitem{Saikumar2017}
N.~Saikumar and S.~HosseinNia, ``Generalized fractional order reset element
  ({GFrORE}),'' in \emph{9th European Nonlinear Dynamics Conference (ENOC)},
  2017.

\bibitem{saikumar2018constant}
N.~Saikumar, R.~K. Sinha, and S.~H. HosseinNia, ``'{Constant in gain Lead in
  phase' element} - application in precision motion control,'' 2018.

\bibitem{arun2018}
A.~Palanikumar, N.~Saikumar, and S.~H. HosseinNia, ``No more differentiator in
  {PID}: Development of nonlinear lead for precision mechatronics,'' pp.
  991--996, June 2018.

\bibitem{saikumar2019resetting}
N.~Saikumar, R.~K. Sinha, and S.~H. HosseinNia, ``Resetting disturbance
  observers with application in compensation of bounded nonlinearities like
  hysteresis in piezo-actuators,'' \emph{Control Engineering Practice},
  vol.~82, pp. 36--49, 2019.

\bibitem{samko1993}
S.~G. Samko, A.~A. Kilbas, and O.~I. Marichev, \emph{Fractional integrals and
  derivatives}.\hskip 1em plus 0.5em minus 0.4em\relax Yverdon: Gordon and
  Breach, 1993.

\bibitem{podlubny1999}
I.~Podlubny, \emph{Fractional differential equations: an introduction to
  fractional derivatives, fractional differential equations, to methods of
  their solution and some of their applications}.\hskip 1em plus 0.5em minus
  0.4em\relax San Diego: Academic Press, 1999.

\bibitem{valerio2013}
D.~Val\'erio and J.~{S\'a da Costa}, \emph{An Introduction to Fractional
  Control}.\hskip 1em plus 0.5em minus 0.4em\relax Stevenage: IET, 2013, iSBN
  978-1-84919-545-4.

\bibitem{Banos2012}
A.~Ba{\~n}os and A.~Barreiro, \emph{Reset control systems}.\hskip 1em plus
  0.5em minus 0.4em\relax Springer Science \& Business Media, 2011.

\bibitem{Guo2009}
Y.~Guo, Y.~Wang, and L.~Xie, ``Frequency-domain properties of reset systems
  with application in hard-disk-drive systems,'' \emph{IEEE Transactions on
  Control Systems Technology}, vol.~17, no.~6, pp. 1446--1453, 2009.

\bibitem{NUIJ20061883}
P.~Nuij, O.~Bosgra, and M.~Steinbuch, ``Higher-order sinusoidal input
  describing functions for the analysis of non-linear systems with harmonic
  responses,'' \emph{Mechanical Systems and Signal Processing}, vol.~20, no.~8,
  pp. 1883--1904, 2006.

\bibitem{Kars}
K.~Heinen, ``Frequency analysis of reset systems containing a {C}legg
  integrator: An introduction to higher order sinusoidal input describing
  functions,'' 2018.

\bibitem{beker2004fundamental}
O.~Beker, C.~Hollot, Y.~Chait, and H.~Han, ``Fundamental properties of reset
  control systems,'' \emph{Automatica}, vol.~40, no.~6, pp. 905--915, 2004.

\bibitem{nevsic2008stability}
D.~Ne{\v{s}}i{\'c}, L.~Zaccarian, and A.~R. Teel, ``Stability properties of
  reset systems,'' \emph{Automatica}, vol.~44, no.~8, pp. 2019--2026, 2008.

\bibitem{lambrechts2005trajectory}
P.~Lambrechts, M.~Boerlage, and M.~Steinbuch, ``Trajectory planning and
  feedforward design for electromechanical motion systems,'' \emph{Control
  Engineering Practice}, vol.~13, no.~2, pp. 145--157, 2005.

\end{thebibliography}

\end{document}